\documentclass[iop,apj,numberedappendix]{emulateapj}
\usepackage{natbib}
\usepackage{graphicx}
\usepackage{amsmath,amsfonts}
\usepackage{multirow}
\usepackage{booktabs}
\usepackage{hyperref}

\slugcomment{Accepted for publication in ApJ, August 26, 2014}

\setcitestyle{notesep={; }}

\def\sgra{\mbox{Sgr A$^{\ast}$}}
\def\lsim{\mathrel{\raise.3ex\hbox{$<$\kern-.75em\lower1ex\hbox{$\sim$}}}}
\def\gsim{\mathrel{\raise.3ex\hbox{$>$\kern-.75em\lower1ex\hbox{$\sim$}}}}
\def\gtwid{\mathrel{\raise.3ex\hbox{$>$\kern-.75em\lower1ex\hbox{$\sim$}}}}
\def\proptwid{\mathrel{\raise.3ex\hbox{$\propto$\kern-.75em\lower1ex\hbox{$\sim$}}}}

\begin{document}
\title{Relative Astrometry of Compact Flaring Structures in Sgr A* with Polarimetric VLBI}
\shorttitle{Polarimetric Relative Astrometry}

\author{Michael D.\ Johnson\altaffilmark{1}, 
Vincent L.\ Fish\altaffilmark{2}, 
Sheperd S.\ Doeleman\altaffilmark{1,2},
Avery E.\ Broderick\altaffilmark{3,4}, 
John F.\ C.\ Wardle\altaffilmark{5}, 
Daniel P.\ Marrone\altaffilmark{6}}
\shortauthors{Johnson et al.}
\altaffiltext{1}{Harvard-Smithsonian Center for Astrophysics, 60 Garden Street, Cambridge, MA 02138, USA}
\altaffiltext{2}{Massachusetts Institute of Technology, Haystack Observatory, Route 40, Westford, MA 01886, USA}
\altaffiltext{3}{Perimeter Institute for Theoretical Physics, 31 Caroline Street North, Waterloo, ON N2L 2Y5, Canada}
\altaffiltext{4}{Department of Physics and Astronomy, University of Waterloo, 200 University Avenue West, Waterloo, ON N2L 3G1, Canada}
\altaffiltext{5}{Department of Physics MS-057, Brandeis University, Waltham, MA 02454-0911}
\altaffiltext{6}{Arizona Radio Observatory, Steward Observatory, University of Arizona, 933 North Cherry Ave., Tucson, AZ 85721-0065, USA}
\email{mjohnson@cfa.harvard.edu} 

\begin{abstract}
We demonstrate that polarimetric interferometry can be used to extract precise spatial information about compact polarized flares of Sgr A*. We show that, for a faint dynamical component, a single interferometric baseline suffices to determine both its polarization and projected displacement from the quiescent intensity centroid.
A second baseline enables two-dimensional reconstruction of the displacement, and additional baselines can self-calibrate using the flare, enhancing synthesis imaging of the quiescent emission. We apply this technique to simulated 1.3-mm wavelength observations of a ``hot spot'' embedded in a radiatively inefficient accretion disk around Sgr A*. Our results indicate that, even with current sensitivities, polarimetric interferometry with the Event Horizon Telescope can achieve ${\sim}5\ \mu{\rm as}$ relative astrometry of compact flaring structures near Sgr A* on timescales of minutes. 
\end{abstract}


\section{Introduction}
\label{sec::Introduction}

The Galactic Center black hole, Sagittarius A* (\sgra), provides an excellent laboratory to study dynamical events in the presence of strong gravity. \sgra\ shows daily flares from radio to X-rays with associated timescales that are comparable to the period of its innermost stable circular orbit, implicating dynamical activity on event-horizon scales \citep{Baganoff_2001,Genzel_2003,Porquet_2003,Eckart_2004,Yusef-Zadeh_2006,Marrone_2006,Marrone_2008}. Yet, the precise nature of these flares is poorly understood.

Very-long-baseline interferometry (VLBI) at 1.3 mm with the Event Horizon Telescope (EHT) has achieved angular resolution for \sgra\ on scales that are commensurate with the apparent size of its photon ring \citep{Doeleman_2008,Fish_2011}. The longest current baseline (Hawaii-Arizona) provides a nominal resolution of $60\ \mu\mathrm{as}$, and the extended array will include baselines with nominal resolutions of ${\sim}20\ \mu\mathrm{as}$ at 1.3 mm, only twice the Schwarzschild radius of \sgra. Also, the development and implementation of ultra-wideband instrumentation now enables high signal-to-noise measurements on timescales of ${\sim}$minutes. For comparison, the period of the innermost stable circular orbit ranges from 30 minutes (for a non-rotating black hole) to 4 minutes (for a maximally-rotating black hole). Hence, by achieving the requisite spatial and temporal resolutions concurrently, the EHT provides a promising avenue to study flares of \sgra. 

Although conventional VLBI employs Earth-rotation synthesis to obtain images, it can also provide detailed information on much shorter timescales. For instance, \citet{Doeleman_HotSpots} demonstrated that analysis of closure amplitudes and phases can readily detect sub-hour periodicities in the Schwarzschild-scale emission structure, such as those from orbiting hot spots embedded in an accretion flow. Similarly,  \citet{Fish_HotSpots} showed that fractional polarization can also sensitively reflect such periodicities. Polarimetry is especially important for sub-mm variability studies of \sgra, which regularly exhibits significant variability in its polarized flux even while the total flux remains relatively constant \citep[e.g.,][]{Marrone_2006}. 

Most efforts to date have focused on observations and interpretations of flares that significantly modulate the total flux of \sgra. Such focus is sensible for near-infrared (NIR) and shorter wavelengths because the flaring emission dominates the energetic output. However, in the sub-mm, most flares only modestly affect the bright quiescent flux. Also, at sub-mm wavelengths, the flaring signatures are both enhanced and persistent in the linear polarization. Indeed, the physical mechanism that gives rise to these flares may well be distinct from that which generates the broadband flares seen up to high energies. Hence, we will emphasize the case of a faint flaring component. 

Aside from their brightness, there are several indications that the flaring structures are extremely compact. Most directly, their variability timescales can be quite short. In addition, fractional polarization of flaring components is high in the mm and NIR \citep{Eckart_2006,Trippe_2007,Marrone_2008,Zamaninasab_2010}, requiring a coherent projected magnetic field direction throughout the flaring region. Finally, flaring structures that lie on Keplerian orbits, such as those arising from magnetic reconnection events or instabilities in the accretion disk, will shear apart on an orbital timescale, although such objects, of characteristic size $\Delta r$, will still remain compact relative to their orbital radius $r$ for $\ll r/\Delta r$ orbital periods. The observed ${\sim}1.5\ \mathrm{hour}$ flare durations necessitate such compactness for this class of flaring models. 

In this paper, we demonstrate that polarimetric VLBI enables relative astrometry of a compact and polarized flaring component. We show that even a \emph{single} baseline with current capabilities can determine the offset of a faint flaring component to ${\sim}5\ \mu\mathrm{as}$, relative to the quiescent intensity centroid and projected along the baseline direction, on timescales of a few minutes or less. The dynamical polarization of the flaring component can be simultaneously measured. Two baselines suffice to reconstruct the two-dimensional trajectory of the flare, and additional baselines provide considerable redundancy, enabling robust inferences. Finally, we demonstrate that the presence of this type of flaring component can act as a phase reference for long baselines, enhancing VLBI imaging capabilities.

\section{Fractional Polarization and VLBI}
\label{sec::FractionalPolarization}

An interferometer samples the two-point function, or \emph{visibility}, of the electric field. For emission from an incoherent source, this function is equal to the Fourier transform of the source brightness distribution \citep[see, e.g.,][]{TMS}:
\begin{align}
\label{eq::vCZ}
\tilde{\mathcal{I}}(\textbf{u}) = \int d^2\textbf{x}\, \mathcal{I}(\textbf{x}) e^{-2\pi i \textbf{u} \cdot \textbf{x}}.
\end{align}
In this expression, $\textbf{u}$ is the projected baseline of the interferometer in wavelengths, and $\textbf{x}$ is an angular coordinate on the sky in radians. An analogous relationship holds for each of the Stokes parameters $\{ \mathcal{I},\, \mathcal{Q},\, \mathcal{U},\, \mathcal{V}\}$, all of which can be sampled in the visibility domain using interferometers that measure the full polarization content of the electric field. Conventional synthesis imaging relies on sufficient sampling of these functions in the visibility plane to reconstruct the Fourier-conjugate images. 

Stations in VLBI are geographically distinct, and cumulative phase errors, principally from the atmosphere, add many unknown turns to the phase of $\tilde{\mathcal{I}}(\textbf{u})$. Nevertheless, intrinsic phase information is still accessible through several pathways. The most common is \emph{closure phase}, the sum of interferometric phase around a closed triangle of baselines, which is immune to station-based phase contributions \citep{Rogers_1974}. 

Fractional polarization can provide identical immunity on individual baselines by phase referencing cross-hand correlations to parallel-hand correlations \citep[see, e.g.,][]{RWB94}. Most VLBI stations record dual circular polarizations (\textbf{L}CP and \textbf{R}CP), so one typical observable is \citep[][Eq.\ 17]{RWB94}
\begin{align}
\label{eq::ExamplePolEq}
\frac{\left \langle R_1 L_2^\ast \right \rangle}{\left \langle R_1 R_2^\ast \right \rangle} \approx g_2 \left[ \breve{m}(\textbf{u}) e^{-2i \phi_2} + D_{1,\rm R} e^{2i \left(\phi_1 - \phi_2 \right)} + D_{2,\rm L}^\ast \right].
\end{align}
Here, the numbered subscript denotes the station, $g_2$ is the quotient of the R and L complex gains of station 2, the $D_{i,x}$ are station- and polarization-dependent ``leakage'' terms, and the $\phi_i$ are the field rotation angles at each station (i.e., the difference between the feed directions and a fixed direction on the sky). The remaining term, $\breve{m}(\textbf{u})$, is the fractional polarization: $\breve{m}(\textbf{u}) \equiv \tilde{\mathcal{P}}(\textbf{u}) / \tilde{\mathcal{I}}(\textbf{u})$, where $\tilde{\mathcal{P}}(\textbf{u}) \equiv \tilde{\mathcal{Q}}(\textbf{u}) + i \tilde{\mathcal{U}}(\textbf{u})$ is the linear polarization in the visibility domain. 
We denote the fractional polarization by a breve rather than a tilde to emphasize that it holds no Fourier relationship with its image-domain analogue $m(\textbf{x}) \equiv \mathcal{P}(\textbf{x})/\mathcal{I}(\textbf{x})$. Indeed, $\breve{m}$ can exhibit many counterintuitive properties. For instance, whereas $|m(\textbf{x})| \leq 1$, $\breve{
m}(\textbf{u})$ can exceed unity, and can even diverge monotonically in all directions.\footnote{As a simple example, suppose that $\mathcal{I}(\textbf{x}) = e^{-|\textbf{x}|^2}$ and $\mathcal{P}(\textbf{x}) = e^{-2|\textbf{x}|^2}$. Then $\breve{m}(\textbf{u}) = \frac{1}{2} \exp\left( \frac{\pi^2}{2} |\textbf{u}|^2 \right)$.} 

In practice, the leakage terms and complex gain quotient tend to be stable over long timescales. Thus, observations of a source with simple or known polarization structure can use the changing field rotation on a long track to identify these instrumental terms. The calibration solution can then be applied to recover the complex fractional polarization of other sources with unknown structure. 

In addition to providing phase information for VLBI, fractional polarization provides several other attractive features. Notably, as Eq.\ \ref{eq::ExamplePolEq} demonstrates, it is only sensitive to the ratio of the orthogonal polarization gains at a station and so tends to be significantly more reliable than absolute flux calibration. Also, as the quotient of two visibilities, fractional polarization is invariant under any convolution in the image domain that affects both circular polarizations equally. Hence, in the noise-free limit, \emph{fractional polarization is immune to scatter-broadening}. 
In general, scattering introduces noise but not bias to fractional polarization measurements.

In the visibility domain, the phase of fractional polarization mixes contributions from the distribution of unpolarized flux, the distribution of polarized flux, and the geometrical structure of the polarization vector field. Nevertheless, as we next demonstrate, we can sometimes use this property to our advantage, recovering astrometric information about the emission components in parallel with inferences about the intrinsic polarization vector field.

Finally, we note an important difference between polarimetric visibilities and those of total flux: the polarization $\tilde{\mathcal{P}}(\textbf{u})$ (and, hence, $\breve{m}(\textbf{u})$) has no conjugation symmetry in the visibility domain. For the VLBI practitioner, this lack of symmetry may be counterintuitive given the familiar symmetry for total flux, $\tilde{\mathcal{I}}(-\textbf{u}) = \tilde{\mathcal{I}}^\ast(\textbf{u})$, a result of the image being real (the relationship holds for each Stokes visibility). An easy way to see that the positive and negative baselines carry distinct polarimetric information is to form the following mixed quantities:
\begin{align}
\frac{1}{2} \left[ \breve{m}(\textbf{u}) + \breve{m}^\ast(-\textbf{u}) \right] &= \frac{\tilde{\mathcal{Q}}(\textbf{u})}{\tilde{\mathcal{I}}(\textbf{u})}\\
\nonumber \frac{1}{2i} \left[ \breve{m}(\textbf{u}) - \breve{m}^\ast(-\textbf{u}) \right] &= \frac{\tilde{\mathcal{U}}(\textbf{u})}{\tilde{\mathcal{I}}(\textbf{u})}.
\end{align}
These identities also demonstrate that, given dual-polarization measurements at each station, the Stokes $\mathcal{Q}$ and $\mathcal{U}$ images can be independently reconstructed \citep[e.g.,][]{Conway_1969}. 

\section{Relative Astrometry Using Fractional Polarization}
\label{sec::Astrometry}

The combination of visibility phase information and immunity to convolution in the image domain makes fractional polarization well suited for relative astrometry. Specifically, an offset in the image domain corresponds to a linear phase gradient with baseline length in the visibility domain, and this phase slope will be identical for emission from point-sources or their smeared counterparts (e.g., from scatter-broadening). This detectable phase gradient can make astrometric information reliable even without full imaging information or specific structural assumptions.

In general, we can decompose the received flux for each Stokes parameter into a sum of contributions from a quiescent (subscript $\mathrm{q}$) and a dynamical (subscript $\mathrm{d}$) component. These quantities are additive in both the image and visibility domains. The measured fractional polarization then takes the form
\begin{align}
\label{eq::QDGeneral}
\breve{m}(\textbf{u},t) &= \frac{ \tilde{\mathcal{P}}_{\rm q}(\textbf{u}) + \tilde{\mathcal{P}}_{\rm d}(\textbf{u},t) }{ \tilde{\mathcal{I}}_{\rm q}(\textbf{u}) + \tilde{\mathcal{I}}_{\rm d}(\textbf{u},t) }.
\end{align}
Of course, even for a constant baseline, this decomposition is not unique, as an arbitrary constant image can be added to the quiescent and subtracted from the dynamical component. One natural assumption to break this degeneracy, in accord with the discussion in \S\ref{sec::Introduction}, is that the dynamical component must be compact relative to the quiescent structure. Hence, we now consider observationally-motivated cases and considerations to illustrate the accessible information for compact flaring regions.

We begin, in \S\ref{sec::PointModel}, with a general description that only assumes that the unscattered dynamical component is completely unresolved by the interferometer. We argue that even a short series of measurements can overconstrain this model, and we explore the solution degeneracies. In \S\ref{sec::ShortBaselines}, we discuss specific considerations for short baselines; we demonstrate that short baselines provide especially robust astrometric inferences, and we show that the astrometric phase can be directly estimated if the dynamical component is faint. In \S\ref{sec::LongBaselines}, we discuss the necessary modifications for longer baselines. We conclude this section by briefly addressing the effects of a changing baseline (\S\ref{sec::ChangingBaseline}), the advantages of known quiescent structure (\S\ref{sec::KnownQuiescent}), and possible extensions to include other VLBI observables (\S\ref{sec::OtherObservables}).

\subsection{A Compact Dynamical Component}
\label{sec::PointModel}

If the dynamical component is sufficiently compact then, in the visibility domain, its baseline dependence is purely an astrometric phase:
\begin{align}
\label{eq::Compact}
\breve{m}(\textbf{u},t) &\approx \frac{ \tilde{\mathcal{P}}_{\rm q}(\textbf{u}) + \mathcal{P}_{\rm d}(t) e^{-2\pi i \textbf{u} \cdot \textbf{x}_{\rm d}(t)} }{ \tilde{\mathcal{I}}_{\rm q}(\textbf{u}) + \mathcal{I}_{\rm d}(t) e^{-2\pi i \textbf{u} \cdot \textbf{x}_{\rm d}(t)}}.
\end{align}
Observe that we have made no assumptions about the quiescent structure. Also, note that we have not specified the origin of the image coordinates, $\textbf{x}$; indeed, Eq.\ \ref{eq::Compact} is invariant under a shift of the origin. However, in \S\ref{sec::ShortBaselines}, we will demonstrate that the centroid of the quiescent flux can be used to define the origin. Also, we again emphasize that this description is unchanged by scatter-broadening of the image, and that the compactness condition will be satisfied whenever the unscattered angular extent of the dynamical component is much smaller than the nominal resolution of the interferometer.

Under this description, the dynamical structure is fully parametrized, on \emph{all} baselines, by three real unknowns ($\textbf{x}_{\rm d}(t)$ and $\mathcal{I}_{\rm d}(t)$) and one complex unknown ($\mathcal{P}_{\rm d}(t)$) at each time. The quiescent structure requires three complex parameters for each baseline, $\tilde{\mathcal{P}}_{\rm q}(\pm\textbf{u})$ and $\tilde{\mathcal{I}}_{\rm q}(\textbf{u})$ (because $\tilde{\mathcal{I}}_{\rm q}(-\textbf{u}) = \tilde{\mathcal{I}}^\ast_{\rm q}(\textbf{u})$). In fractional polarization alone, each baseline provides two independent complex measurements, $\breve{m}(\pm\textbf{u})$. Because many mm-VLBI sites are phased arrays, which give the zero-baseline properties in parallel with VLBI, when we refer to single-baseline conclusions, we will implicitly assume concurrent zero-baseline measurements, which provides another complex measurement $\breve{m}(\textbf{0})$. Hence, even for a short time series of measurements, a \emph{single} baseline suffices to overconstrain 
the system, which is then sensitive to the 
projected displacement of the flare along the baseline direction.

At this point, we can also understand the potential solution degeneracies. First, as with other VLBI observables, fractional polarization is insensitive to the image centroid; yet, we will show that one can determine the offset of a flare relative to the centroid of the quiescent flux (see \S\ref{sec::ShortBaselines}). Also, measurements are only sensitive to the fractional polarization of the flare down-weighted by its relative intensity. And, with a single baseline, if the phase of $\tilde{\mathcal{I}}_{\rm q}$ is unknown, then there is an unknown, constant offset of the inferred position (see \S\ref{sec::LongBaselines}). 

However, perhaps the most important solution degeneracy arises from wraps in the astrometric phase. Of course, each baseline is only sensitive to the projected flux along the baseline direction (a corollary of the projection-slice theorem). In addition, the fractional polarization is invariant under integer shifts of $\textbf{u} \cdot \textbf{x}_{\rm d}(t)$; any inferred offset is only determined up to integer multiples of the nominal resolution of the baseline ($\lambda /D$) along the baseline direction. Thus, a single baseline can estimate the projected displacement of a flare along the baseline direction, modulo $\lambda/D$. 
A second, non-parallel baseline limits the vector offset to a lattice of points that are determined by linear combinations with integer weights of the two vectors $\textbf{u}_1^{\perp}/\left(\textbf{u}_1^{\perp} \cdot \textbf{u}_2  \right)$ and $\textbf{u}_2^{\perp}/\left(\textbf{u}_2^{\perp} \cdot \textbf{u}_1 \right)$, where the $\textbf{u}_i$ are the baseline vectors and $\textbf{u}_i^{\perp} \cdot \textbf{u}_i = 0$. Interestingly, a third baseline that completes a triangle of stations, $\textbf{u}_3 = \textbf{u}_2 - \textbf{u}_1$, does not reduce this lattice because any lattice point $\textbf{x}$ has $\textbf{u}_3 \cdot \textbf{x} = \left(\textbf{u}_2 - \textbf{u}_1 \right) \cdot \textbf{x} \in \mathbb{Z}$. Hence, four stations are required to soften the phase-wrap degeneracy.

\subsubsection{Considerations for Short Baselines}
\label{sec::ShortBaselines}

For astrometric purposes, short baselines provide two advantageous features. The first is that the $\lambda/D$ offset ambiguity is less problematic. The second, and most important, benefit is that sufficiently short baselines do not significantly resolve structure in the quiescent flux associated with \sgra, allowing them to use the quiescent emission centroid to define the image center. Explicitly, 
\begin{align}
\label{eq::Centroid}
\tilde{\mathcal{I}}_{\rm q}(\textbf{u}) &= \sum_{n=0}^\infty \frac{1}{n!} \left( \textbf{u} \cdot \nabla \right)^n \tilde{\mathcal{I}}_{\rm q}(\textbf{0})\\
\nonumber &= \sum_{n=0}^\infty \frac{(-2\pi i)^n}{n!} \int d^2 \textbf{x}\, \left( \textbf{u} \cdot \textbf{x} \right)^n {\mathcal{I}}_{\rm q}(\textbf{x})\\
\nonumber &\approx \left \langle \mathcal{I}_{\rm q}(\textbf{x}) \right \rangle -2\pi i \textbf{u} \cdot \left \langle \textbf{x}\, \mathcal{I}_{\rm q}(\textbf{x}) \right \rangle.
\end{align}
This representation also follows from expanding the exponential in the Fourier relationship between the image and visibility domains (Eq.\ \ref{eq::vCZ}). 
Because $\left \langle \mathcal{I}_{\rm q}(\textbf{x}) \right \rangle \in \mathbb{R}^+$, the leading order change on a short baseline is an imaginary part that is proportional to the centroid of the quiescent flux projected along the baseline direction. So by taking $\tilde{\mathcal{I}}_{\rm q}(\textbf{u})\in \mathbb{R}^+$ on sufficiently short baselines, all inferred astrometric displacements will be relative to the centroid of the quiescent flux. The fall of flux with baseline (i.e., the second-order correction of Eq.\ \ref{eq::Centroid}) assesses the validity of this approximation, as the next contribution to phase arises at third order.

Short baselines also provide a regime in which variations may be significantly stronger in polarization than in total flux, a consequence of low image-averaged polarization. For \sgra\ at mm wavelengths, for instance, the $\lsim 10\%$ polarization seen on short baselines implies that the fractional variations induced by a flare can be an order of magnitude stronger in polarization than in total flux. 

It is instructive to consider a limiting case: short-baseline observations of a faint, highly-polarized flare. If the baseline is sufficiently short, then we can linearize $\breve{m}(\textbf{u},t)$ in baseline. Neglecting the contribution of the flare to total flux leaves $\tilde{\mathcal{I}}(\textbf{u})$ constant to leading order (see Eq.\ \ref{eq::Centroid}) so that
\begin{align}
\label{eq::ShortBaselineExpand}
\breve{m}(\textbf{u},t) &\approx \frac{ \tilde{\mathcal{P}}_{\rm q}(\textbf{0}) + \textbf{u} \cdot \nabla \tilde{\mathcal{P}}_{\rm q}(\textbf{0}) + \mathcal{P}_{\rm d}(t) e^{-2\pi i \textbf{u} \cdot \textbf{x}_{\rm d}(t)}}{ \tilde{\mathcal{I}}_{\rm q}(\textbf{0})}.
\end{align}
We have not linearized the astrometric phase term, which may show variations on baselines that are insensitive to quiescent structure. 
Also, without loss of generality, we can simply normalize the integrated flux $\tilde{\mathcal{I}}_{\rm q}(\textbf{0})$ to be unity. 

We now use this approximation to motivate that the astrometric component of the flare can be robustly isolated using even a single baseline. For example, observe that
\begin{align}
\label{eq::ratiotest}
\frac{\breve{m}(-\textbf{u},t) - \breve{m}(\textbf{0},t) + \textbf{u} \cdot \nabla \tilde{\mathcal{P}}_{\rm q}(\textbf{0})}{\breve{m}(\textbf{u},t) - \breve{m}(\textbf{0},t) - \textbf{u} \cdot \nabla \tilde{\mathcal{P}}_{\rm q}(\textbf{0})} &\approx -e^{2\pi i \textbf{u} \cdot \textbf{x}_{\rm d}(t)} .
\end{align}
The restrictiveness of this relationship (i.e., a norm of unity) demonstrates that the system is overconstrained, provided that the dynamical polarization and astrometric phase are non-zero, so the single undetermined parameter, $\nabla \tilde{\mathcal{P}}_{\rm q}(\textbf{0})$, is accessible from a series of measurements (in practice, through non-linear model fitting). After this determination, Eq.\ \ref{eq::ratiotest} fully decouples the astrometric phase of the flare.

\subsubsection{Modifications for Intermediate to Long Baselines}
\label{sec::LongBaselines}

Longer baselines present several advantages and slightly different challenges. 
The influence of the compact flaring component will become increasingly dominant as more of the comparatively extended quiescent structure is resolved. 
But the most important modification on longer baselines is that the geometrical distribution of flux can introduce a visibility phase, so even for a faint flaring component, imposing $\tilde{\mathcal{I}}_{\rm q}(\textbf{u}) \in \mathbb{R}^+$ is no longer equivalent to phase referencing to the centroid of the quiescent flux. In the most general treatment, this new unknown phase is entirely degenerate with the astrometric phase and introduces an unknown offset. Given redundancy from multiple baselines, this property is advantageous. That is, if shorter baselines determine the offset and polarization of the flaring structure, then longer baselines can phase reference using that information --- a form of self-calibration \citep{TMS}. 

As a simple proof-of-concept, consider a series of measurements over which the baseline, as well as the flaring flux and polarization are constant, but with significant variation in the astrometric phase $\theta(t) \equiv -2\pi \textbf{u} \cdot \textbf{x}_{\rm d}(t)$, which we assume is known from inferences using shorter baselines. Then, a long baseline measures
\begin{align}
\breve{m}(\textbf{u},t) &\approx \frac{ \tilde{\mathcal{P}}_{\rm q} + \mathcal{P}_{\rm d} e^{i \theta(t)} }{ \tilde{\mathcal{I}}_{\rm q} + \mathcal{I}_{\rm d} e^{i \theta(t)}}.
\end{align}
In this case, the average over $\theta$ gives the quiescent fractional polarization, $\breve{m}_{\rm q} \equiv \tilde{\mathcal{P}}_{\rm q} / \tilde{\mathcal{I}}_{\rm q}$, and the first Fourier coefficient (i.e., the average when weighted by $e^{-i\theta}$) gives $\left( \tilde{\mathcal{I}}_{\rm q} \mathcal{P}_{\rm d} - \mathcal{I}_{\rm d} \tilde{\mathcal{P}}_{\rm q} \right)  / \tilde{\mathcal{I}}_{\rm q}^2 = \left( \mathcal{P}_{\rm d} - \mathcal{I}_{\rm d} \breve{m}_{\rm q} \right)/\tilde{\mathcal{I}}_{\rm q}$. This combination, in addition to inferred dynamical characteristics using shorter baselines, immediately gives the phase of $\tilde{\mathcal{I}}_{\rm q}$ on the long baseline. 

Thus, on longer baselines we have the somewhat paradoxical situation in which the presence of source variability can \emph{enhance} measurements of the quiescent structure.

\begin{figure*}[t]
\includegraphics*[width=\textwidth]{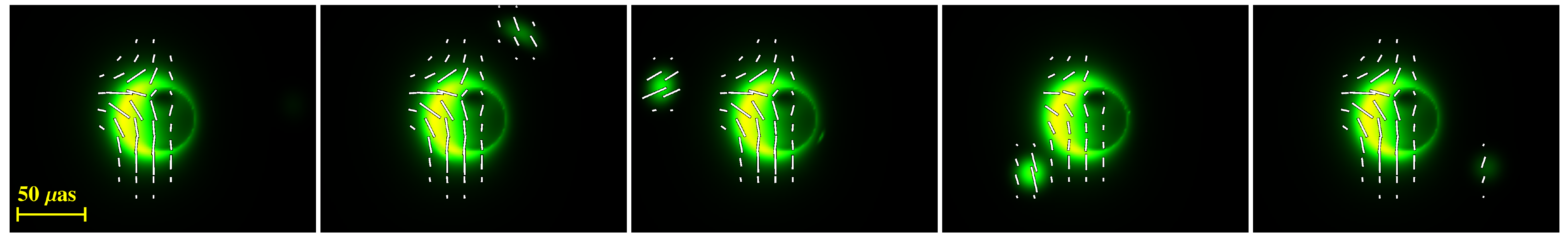}
\caption{
Unscattered ray-traced intensity images of our simulated hot spot model with polarization ticks overlaid. Plotted frames correspond to orbital phases of 0.0, 0.2, 0.4, 0.6, and 0.8. The central, quiescent emission exhibits a bright circular feature of ${\sim} 50\ \mu{\rm as}$ diameter corresponding to the lensed photon ring of the black hole, which surrounds the black hole ``shadow'' \citep{Falcke_2000}. For comparison, the Schwarzschild radius subtends ${\sim} 10\ \mu{\rm as}$. 
The crescent morphology arises from Doppler boosting/deboosting on the approaching/receding side of the accretion disk; these effects are also apparent in the varying brightness of the hot spot throughout its orbit. 
}
\label{fig::SampleFrames}
\end{figure*}

\subsection{Effects of Changing Baseline}
\label{sec::ChangingBaseline}

In practice, each VLBI baseline varies with time. Indeed, secular changes in the quiescent structure are methodologically equivalent to variations with baseline. However, flares of \sgra\ typically vary on timescales of minutes while baselines vary on timescales of hours. Nevertheless, we briefly consider several strategies to gracefully incorporate baseline variation into the preceeding discussion.  

Because the baseline variation is known, the changing astrometric phase, $\textbf{u}(t) \cdot \textbf{x}_{\rm d}(t)$, can easily be incorporated in full generality. Hence, only the changing quiescent properties with baseline, $\tilde{\mathcal{I}}_{\rm q}$ and $\tilde{\mathcal{P}}_{\rm q}$, must be estimated. 

For measurements of a faint flare on a single baseline, recall that the fractional polarization is only sensitive to the dynamical polarization, downweighted by its relative flux. Hence, we may absorb overall variations of $\tilde{\mathcal{I}}_{\rm q}$ with baseline into the dynamical polarization $\tilde{\mathcal{P}}_{\rm d}$, and we only must account for variations in $\tilde{\mathcal{P}}_{\rm q} / \tilde{\mathcal{I}}_{\rm q}$. However, the changing projection direction $\hat{\textbf{u}}$ for the astrometric offset must be considered when interpreting the resulting inferences. 

More generally, because a small number of measurements suffice to decouple the quiescent and dynamical parameters, one can partially account for a changing baseline by subdividing the data into short intervals and determining solutions for each independently. We demonstrate the efficacy of this simple approach in \S\ref{sec::Fitting}. More sophisticated methods include constructing a model image of the quiescent structure or linearizing the variations in differential baseline or time (since variations with baseline are monotonic). The latter is particularly effective for short baselines (see, e.g., Eq.\ \ref{eq::ShortBaselineExpand}).

\subsection{Advantages of Known Quiescent Structure}
\label{sec::KnownQuiescent}

Prior knowledge of the quiescent structure would significantly strengthen this technique. Indeed, there are several indications that the quiescent structure of \sgra\ may be stable over long timescales or, at least, may present stable features. For example, in the total sub-mm flux, there is evidence for structural consistency during bright flares \citep{Fish_2011}. Also, the linear polarization magnitude and direction \citep{Bower_2005,Marrone_2007} and the circular polarization handedness \citep{Bower_2002,Munoz_2012} have shown long-term stability. In addition, simulated images of the polarized emission predict smooth and stable features related to order in the magnetic field and accretion flow geometry \citep{Bromley_2001,Broderick_Blandford_2004,Broderick_Loeb_2006,Huang_2008,Shcherbakov_2012}. Even in lieu of long-term stability, if the onset of a flare is sudden, then data preceding or following the flare may suffice to characterize the quiescent structure during the flare. 

If the quiescent structure is known and the flare contributes negligible total flux, then the downweighted polarization can be determined immediately, and the $\breve{m}(\pm \textbf{u})$ comparison directly yields the astrometric phase (see Eq.\ \ref{eq::Compact}). Even if the flare is not faint, the inferences are significantly tightened because we have made no assumptions (e.g., continuity) about how the dynamical component evolves in time. Thus, the set of equations that describe the observed fractional polarization (Eq.\ \ref{eq::QDGeneral}) only couple through the quiescent terms, $\tilde{\mathcal{P}}_{\rm q}(\textbf{u})$ and $\tilde{\mathcal{I}}_{\rm q}(\textbf{u})$. In fact, knowing only the zero-baseline quiescent terms will significantly improve fits and eliminate many solution degeneracies for complex models.

\subsection{Inclusion of Other VLBI Observables}
\label{sec::OtherObservables}

Thus far, we have only considered VLBI measurements of fractional polarization, although one can easily incorporate measurements of the total flux $\left| \tilde{\mathcal{I}}(\textbf{u},t) \right| = \left| \tilde{\mathcal{I}}_{\rm q}(\textbf{u}) + \tilde{\mathcal{I}}_{\rm d}(\textbf{u},t) \right|$ and closure phase. Note that, because our treatment assumes a point-like flare, these flux measurements correspond to those of the {unscattered} image --- scattering effects must either be weak or known and inverted (by dividing each visibility measurement by the Fourier transform of the scattering kernel on the same baseline) \citep{Fish_Deblurring}. For Sgr A* at 1.3 mm, the latter option is possible by extrapolating the longer-wavelength parameters and scaling of the scattering \citep{Shen_2005,Bower_2006,Lu_2011}.

\section{Application to Simulated Data}
\label{sec::Simulations}

We now apply the ideas of \S\ref{sec::Astrometry} to analyze simulated data of a ``hot spot'' orbiting \sgra.  We emphasize that these simulated data, although motivated and guided by observations, simply provide a controlled setting in which we can test critical assumptions that underlie the above analysis and do not represent the full intended scope of applicability or even a preferred physical description of the sub-mm flares.

\subsection{Model Details}

For brevity, we will only summarize the salient features and limitations of this simulation, which was originally presented in \citet{Doeleman_HotSpots} and \citet{Fish_HotSpots} (their ``Model F'') and was based on the work of \citet{Broderick_Loeb_2006}. We refer the reader to those original works for a complete discussion.

In this simulation, the quiescent emission arises from a radiatively inefficient, Keplerian accretion disk threaded by a toroidal magnetic field. The hot spot is modeled as a compact, Gaussian overdensity of non-thermal electrons that is frozen into the accretion flow. The black hole has zero spin; the accretion disk is inclined $60^\circ$ relative to the line of sight with a major-axis position angle that is $90^\circ$ east of north. The hot spot orbits with a period of 167 minutes (the innermost stable circular orbit has a period of 27 minutes). The quiescent and hot spot emission is primarily synchrotron. 

Images with pixel resolutions of $\left(1.46\ \mu{\rm as}\right)^2$ were computed at 100 equally spaced positions within a single orbit using a fully relativistic ray-tracing code with polarized radiative transfer (see Figure \ref{fig::SampleFrames} for examples). Although irrelevant to our astrometric demonstration, a major limitation of these simulations is that they do not account for evolution (e.g., shearing or cooling) of the hot spot throughout its orbit.

\subsection{Generation of Synthetic VLBI Data}

\begin{figure}[b]
\centering
\includegraphics[width=0.43\textwidth]{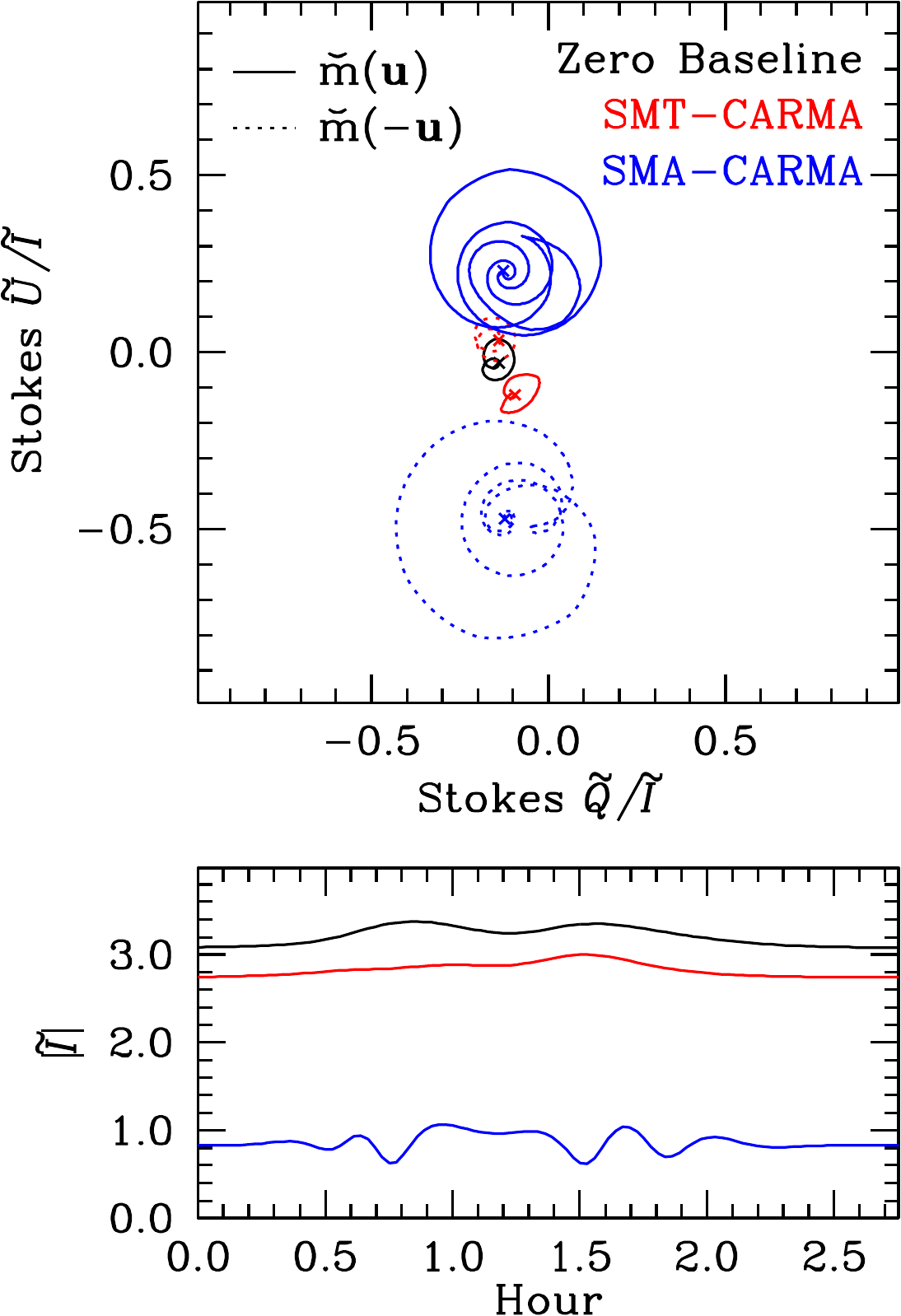}
\caption{
Complex fractional polarization (\textit{upper}) and total flux (\textit{lower}) in the visibility domain over the full hot-spot period for three fixed baselines. The hot spot contributes negligible flux near an orbital phase of 0.0 (marked by a cross in the upper figure), as Figure \ref{fig::SampleFrames} indicates, and a maximum of 10\% of the quiescent flux. On the long baseline (SMA-CARMA), the polarization trace undergoes many loops as the astrometric phase wraps through many turns. Note that because the baselines are fixed, all variation is from temporal changes in the source.  
}
\label{fig::PolTrace}
\end{figure}

We generated synthetic VLBI data (see Figure \ref{fig::PolTrace}) for the ray-traced model images using an array that reflects the locations and specifications of three current 1.3-mm VLBI sites. These included the Combined Array for Research in Millimeter-wave Astronomy (CARMA) in California, the Submillimeter Telescope (SMT) in Arizona, and the Submillimeter Array (SMA) on Mauna Kea in Hawaii. The SMA does not currently have dual-polarization receivers, although simultaneous observations with the James Clerk Maxwell Telescope (JCMT) enable a single effective dual-polarization station in Hawaii. For simplicity, we simulated dual-polarization observations at each site. For telescope parameters, including expected VLBI performance, see \citet{Doeleman_HotSpots} or \citet{Lu_2014}.

\subsection{Noise in the Simulated Data}
\label{sec::Noise}
Two types of errors limit polarimetric VLBI observations. The first, thermal noise, decreases with the square root of observing bandwidth and integration time, effectively setting the time resolution of measurements. The second, instrumental ``leakage'' (see Eq.\ \ref{eq::ExamplePolEq}), is largely unaffected by increased integration time or bandwidth. However, leakage errors are invertible and can be determined using measurements of bright calibrators with simple structure; the residual uncertainty is typically ${\ll}1\%$ \citep{RWB94}. Although leakage terms tend to be stable over long timescales (months to years), their influence on the data change over ${\sim}$hour timescales as the telescope feeds rotate with respect to a fixed direction on the sky. Hence, residual calibration errors will slightly bias the baseline-dependent {quiescent} polarization estimates ($\mathcal{P}_{\rm q}(\textbf{u})$). Since we are principally concerned with astrometry of the dynamical component, we only include thermal noise. 
We used the system equivalent flux densities (SEFDs) reported in \citet{Lu_2014} with 4-GHz observing bandwidths and 100-second integration times. 

As we have discussed in \S\ref{sec::FractionalPolarization} and \S\ref{sec::OtherObservables}, the effect of scattering is to increase the noise in these simulations. At \mbox{1.3 mm}, the diffractive scale of the scattering is thousands of kilometers \citep{Shen_2005,Lu_2011}. Thus, on the ${\sim} 900\ {\rm km}$ SMT-CARMA baseline (used in the following \S\ref{sec::Fitting}), the scattering will only decrease the signal-to-noise by a few percent. Indeed, the signal-to-noise is never reduced by more than a factor of ${\sim}5$ over the \emph{entire} range of current and projected EHT baselines \citep{Fish_Deblurring}.

\subsection{Fitting Procedure and Results}
\label{sec::Fitting}

At 1.3 mm, the shortest EHT baseline (other than array measurements) is SMT-CARMA ($\lambda/D \sim 300\ \mu\mathrm{as}$). This baseline provides an excellent application of the ``short baseline'' assumptions outlined in \S\ref{sec::ShortBaselines}. Specifically, we assume that $\tilde{\mathcal{I}}_{\rm q} \in \mathbb{R}^+$, although we do not linearize the quiescent structure in baseline (Eq.\ \ref{eq::ShortBaselineExpand}). 

We sampled model images on a fixed SMT-CARMA baseline over the full orbital period and then fit the synthetic data in two ways. We first fit the synthetic data using only fractional polarization $\breve{m}(\textbf{u})$ measurements; in this case, we also assumed that the contribution of the flaring flux was negligible. Thus, the model was given by Eq.\ \ref{eq::Compact} with $\mathcal{I}_{\rm d}(t) \rightarrow 0$. We then fit the synthetic data using both fractional polarization and total flux $\left| \tilde{\mathcal{I}}(\textbf{u}) \right|$ measurements. In this case, we did not neglect the dynamical contribution to flux.   

The full set of independent model parameters was then $\{ \tilde{\mathcal{I}}_{\rm q}(\textbf{0}),\, \tilde{\mathcal{I}}_{\rm q}(\textbf{u}),\, \tilde{\mathcal{P}}_{\rm q}(\pm \textbf{u}) \}$ to describe the quiescent structure, and $\{ \mathcal{P}_{\rm d}(t),\, \textbf{u}\cdot \textbf{x}_{\rm d}(t) \}$ at each time step. For fits with $\left| \tilde{\mathcal{I}}(\textbf{u}) \right|$, there was an additional parameter $\mathcal{I}_{\rm d}(t) \in \mathbb{R}$ at each time step, while for fits without $\left| \tilde{\mathcal{I}}(\textbf{u}) \right|$, the flux scaling is arbitrary and so we simply took $\tilde{\mathcal{I}}_{\rm q}(\textbf{0}) \equiv 1$ without loss of generality. We made no assumptions about the relationship between the time-variable parameters at different time steps (e.g., continuity). 

In each of these cases, we fit the data both with and without thermal noise, to assess systematic errors in our model and the importance of thermal noise. Our fits used the standard weighted-least-squares prescription. For the noise-free case, we weighted all residuals (data minus model) equally, while for the data with thermal noise, we normalized residuals by the standard deviation of the thermal noise. As mentioned in \S\ref{sec::KnownQuiescent}, the fitted parameters at different time steps only couple through the parameters that characterize the quiescent emission. Because of this property, even for a complicated $\chi^2$ hypersurface, finding the global maximum and its associated uncertainties is straightforward by an exhaustive search over the small number of quiescent parameters.

\begin{figure}[t]
\includegraphics*[width=0.475\textwidth]{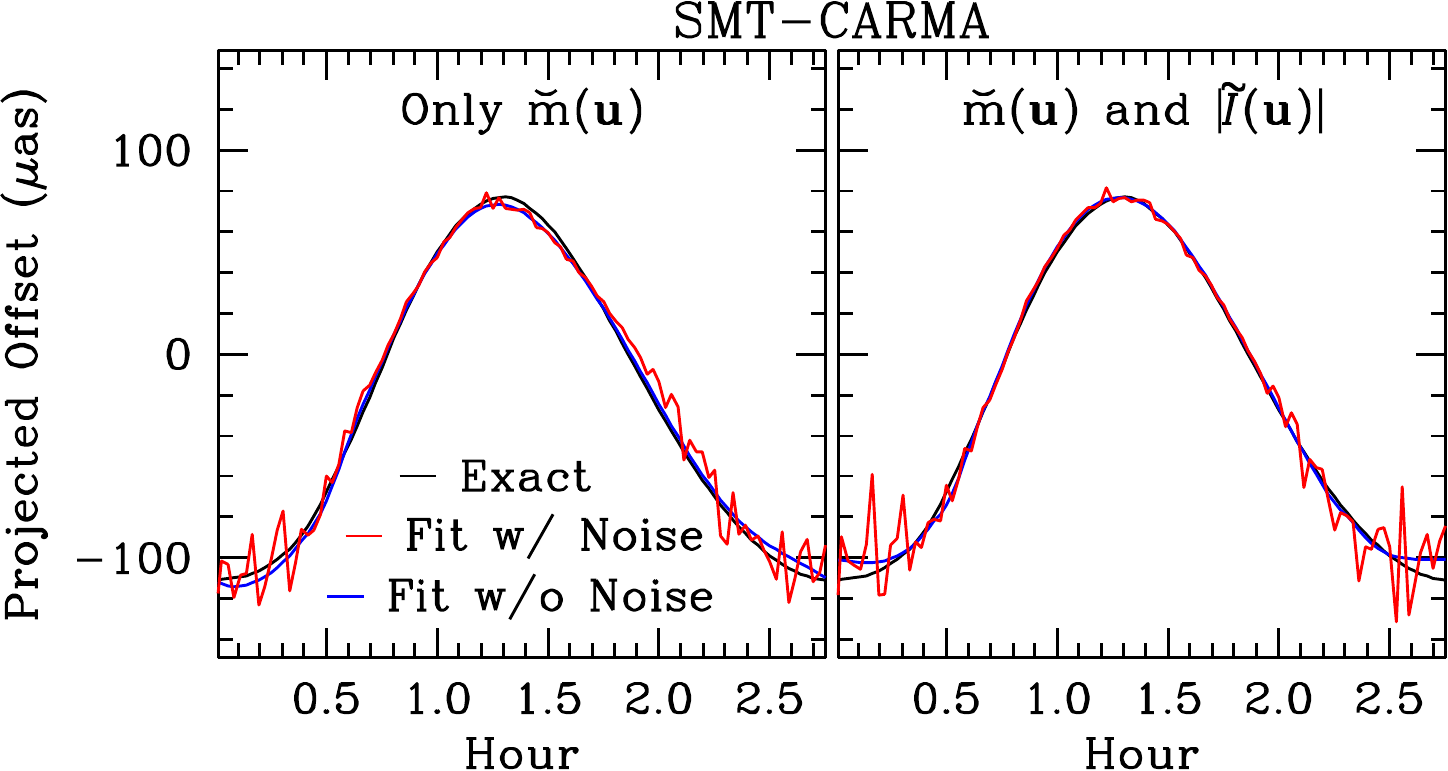}
\caption{
Exact and inferred projected displacement of the hot spot as a function of time for fits to synthetic data with and without thermal noise on the SMT-CARMA baseline (see Figure~\ref{fig::PolTrace}). The fitting details and procedure are given in \S\ref{sec::Fitting}. Because the fit with fractional polarization measurements alone (\emph{left}) assumed that the flaring flux was negligible, the inclusion of estimated flux density $\left| \tilde{\mathcal{I}} \right|$ (\emph{right}) improves the inferences slightly in the central orbital phases, where the hot spot is brightest. Over the majority of the orbital period, 
when the hot spot is not extremely Doppler deboosted, 
the systematic errors in both fits are within $5\ \mu\mathrm{as}$. For this baseline, the offset ambiguity is $\lambda/D \approx 330\ \mu\mathrm{as}$.  
}
\label{fig_1DAstrometry}
\end{figure}

Figure \ref{fig_1DAstrometry} compares the inferred projected displacement of the hot spot with time to the exact values, calculated using the model images. 
These results demonstrate that, despite its modest resolution and comparatively high SEFD, the SMT-CARMA baseline can yield excellent astrometric inferences, even for faint flares. Our simplified model assumptions do not generate large systematic errors, and we can readily recover detailed orbital information, even with expected levels of thermal noise.

Next, we repeated these steps but included the effect of Earth rotation so that the baselines gradually changed with time. This change affects the projected direction of the offset and the translation between astrometric phase and displacement. Most importantly, this change introduces slow, secular variations in the quiescent parameters, $\tilde{\mathcal{I}}_{\rm q}(\textbf{u})$ and $\tilde{\mathcal{P}}_{\rm q}(\textbf{u})$ associated with each baseline. However, as discussed in \S\ref{sec::PointModel}, the set of equations is significantly overconstrained after a short series of measurements, and the hot spot parameters change much more rapidly than those of the quiescent emission. Thus, we subdivided the period into five equal segments and found independent solutions for each assuming a constant baseline. 

\begin{figure}[t]
\includegraphics*[width=0.475\textwidth]{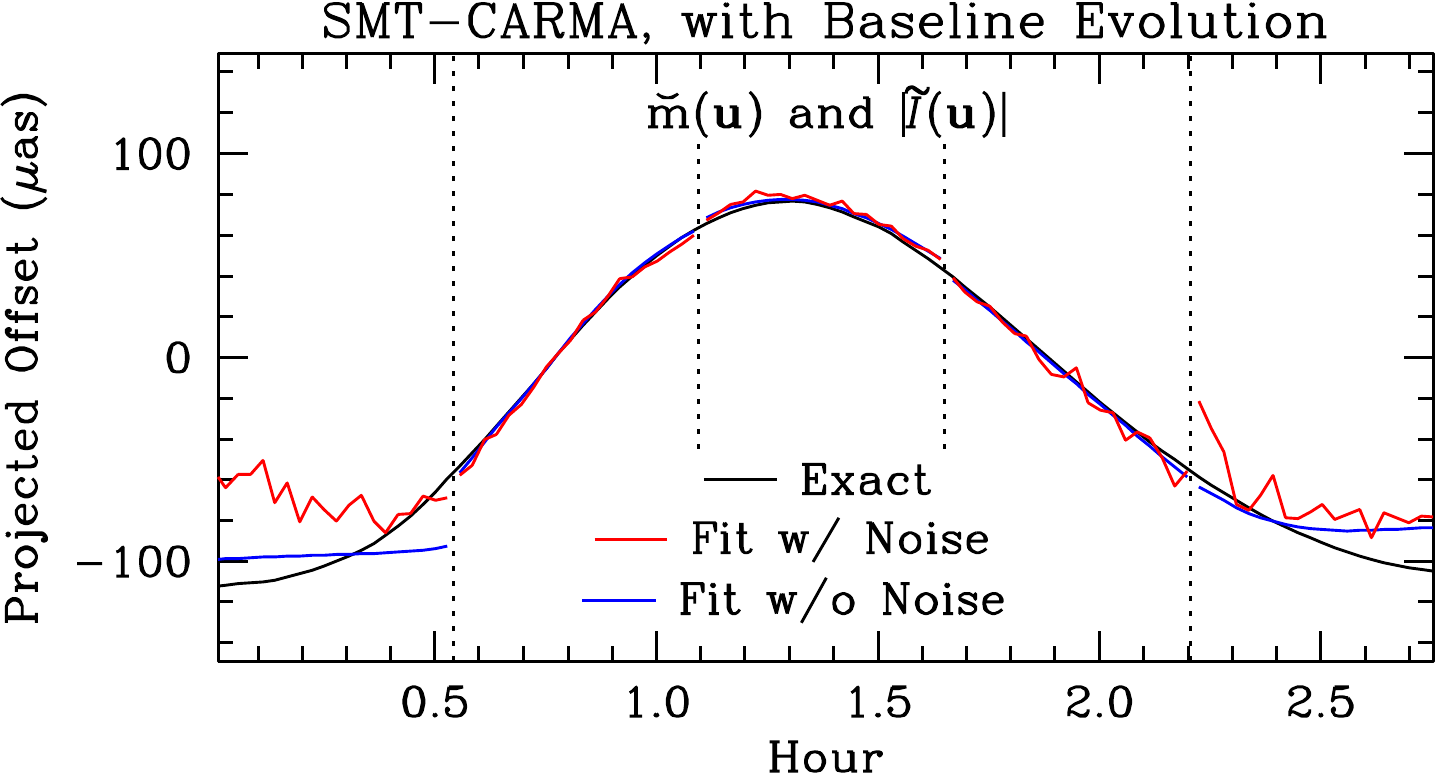}
\caption{
Same as Figure \ref{fig_1DAstrometry}, but for synthetic data that now include baseline evolution in time.
To account for the slow changes in the quiescent parameters, we divided the data into five equal ${\sim}33$-minute segments (separated by dashed vertical lines), each with 20 measurements, and then we derived an independent solution for each segment. 
As in Figure \ref{fig_1DAstrometry}, the solutions are unreliable near an orbital phase of zero because the hot spot contributes almost no flux there ($\lsim 2\%$ of the quiescent flux), but the bias throughout the remaining part of the orbit is a few microarcseconds at most. The fits are poorer than those of Figure \ref{fig_1DAstrometry} when the hot spot is faint because they no longer utilize the well-determined quiescent parameters from the middle of the orbit.
}
\label{fig_1DAstrometry_wBaseline}
\end{figure}

The results, shown in Figure \ref{fig_1DAstrometry_wBaseline}, are less reliable when the hot spot is extremely faint, but the residual bias incurred by the baseline evolution is only a few microarcseconds when the hot spot contributes more than a few percent of the total flux. This example demonstrates that our method is insensitive to slow changes in the quiescent parameters, either from processes intrinsic to the source or from observational considerations, such as a changing baseline or calibration errors. 

\section{Discussion}
\label{sec::Discussion}

\subsection{Comparisons with \textit{GRAVITY}}
The possibility of microarcsecond astrometry with the EHT nicely complements the similar efforts using the \textit{GRAVITY} instrument at the Very Large Telescope Interferometer (VLTI) \citep{Eisenhauer_2008,Bartko_2009,Hamaus_2009,Eckart_2010,Vincent_2011}. Operating at $\lambda \sim 2\ \mu{\rm m}$ with a target accuracy of $10\ \mu\mathrm{as}$ for relative astrometry, \textit{GRAVITY} will identify subtle changes in the centroid of the total flux emission from \sgra. Because the quiescent emission from \sgra\ is negligible in the NIR, this centroid motion is expected to be dominated by flaring structures. 

The redundancy and additional spatial information of 1.3-mm VLBI with the EHT can considerably extend these studies. In particular, VLBI can provide information about the compactness and anisotropy of the flaring structures in addition to the centroid motion. Also, mm-wavelength studies benefit from a ${\sim}25\times$ longer synchrotron cooling time ($\propto \sqrt{\lambda}$) \citep{Broderick_Loeb_2006}, which may enable the tracking of features over many orbits, especially with the addition of southern VLBI sites (ALMA and the South Pole Telescope) that can observe \sgra\ for many hours uninterrupted. Moreover, because the persistence of NIR flares beyond their cooling time implies continuing particle acceleration, concurrent astrometric information at this pair of frequencies separated by almost three orders of magnitude will provide significant insight into the formation and dissipation of flares and the sites of particle acceleration.

\subsection{Summary}

We have demonstrated that polarimetric mm-VLBI can determine the relative displacement of a differentially polarized, compact flaring component in \sgra\ to ${\sim}\mu\mathrm{as}$ accuracy on $\sim$minute timescales --- resolutions that are well-matched to the gravitational radius and timescales of \sgra. 

In short, our technique leverages the phase information afforded by fractional polarization to perform astrometry of a flaring component relative to the centroid of the quiescent flux. Perhaps the most surprising result is that a \emph{single} short baseline can achieve reliable relative astrometry without requiring a quiescent source model. This result arises because astrometric phase effects are dominant on short baselines and can be significant even when the change in flux is mild. 
The particular effectiveness of polarimetry, even with a small number of baselines, is a consequence of polarization providing two independent measurements, $\left \langle R_1 L_2^\ast \right \rangle$ and $\left \langle L_1 R_2^\ast \right \rangle$, on each baseline (see \S\ref{sec::FractionalPolarization}).

The major benefits of our proposed technique are the following:
\begin{enumerate}
\item Our approach requires no assumptions about the structure of the quiescent emission or properties of the position of the flare (e.g., periodicity).
\item We only require that the flaring emission be polarized and compact --- features with observational support. In the particular case of \sgra, our technique is applicable to a broad class of dynamical activity, including local hot spots in an accretion flow, shocks propagating down a jet, or even an orbiting pulsar or magnetar. 
\item Even a single baseline suffices to overconstrain the system, thereby determining whether the flaring structure is compact in addition to its dynamical properties. Two baselines can reconstruct the two-dimensional apparent displacement, and additional baselines provide considerable redundancy, potentially enabling measurements of visibility phase on long baselines. Each solution also estimates the changing polarization of the flare.
\item Our method is insensitive to propagation effects, such as scatter broadening or a rotation measure gradient across the image.
\end{enumerate}

We have shown that the shortest baseline (SMT-CARMA) of the Event Horizon Telescope can readily achieve ${\sim}\mu{\rm as}$ relative astrometry of a compact flaring component. 
Planned extensions of the EHT to include the Large Millimeter Telescope (LMT) on Sierra Negra in Mexico, the Atacama Large Millimeter/submillimeter Array (ALMA) on the Chajnantor plateau in Chile, and the South Pole Telescope in 2015 will enable yet more stringent constraints. 
One challenge for the EHT is that the smallest triangle of stations with mutual visibility, CARMA-SMT-LMT, is nearly collinear and, therefore, poorly suited for two-dimensional reconstructions. The baseline from the 30-m Institut de Radioastronomie Millim\'{e}trique (IRAM) telescope on Pico Veleta to the IRAM Plateau de Bure Interferometer will provide a short, nearly orthogonal baseline (albeit not simultaneously), potentially aiding this ambiguity if the flares occur predominantly on a single plane. Other triangles, such as SMA-SMT-CARMA, permit two-dimensional reconstructions subject to a phase wrap degeneracy, and the inclusion of a fourth site will effectively eliminate this degeneracy (see \S\ref{sec::PointModel}).

Our treatment, which primarily serves as a proof-of-concept, can be expanded in several directions. With additional interferometric observables (such as closure phases and amplitudes) and the redundancy of multiple baselines, one could relax the model assumptions outlined in \S\ref{sec::Astrometry}. Extensions could include the contribution of the lensed secondary image, for instance. And, although we have assumed a monochromatic description, planned dual-sideband and 0.87 mm capabilities for the EHT could identify the Faraday rotation of the flaring component, enabling a dynamical tomographic probe of magnetic field strength.

\acknowledgments
We thank the National Science Foundation (AST-1207752, AST-1310896, and AST-1211539) and the Gordon and Betty Moore Foundation (\#GBMF-3561) for financial support of this work. A.E.B. receives financial support from the Perimeter Institute for Theoretical Physics and the Natural Sciences and Engineering Research Council of Canada through a Discovery Grant. Research at Perimeter Institute is supported by the Government of Canada through Industry Canada and by the Province of Ontario through the Ministry of Research and Innovation.

\bibliography{Polarimetric_Astrometry.bib}

\end{document}